# Molecular-Scale Insights into the Heterogeneous Interactions Between an *m*-Terphenyl Isocyanide Ligand and Noble Metal Nanoparticles

Liya Bi[1,2,‡], Yufei Wang[2,3,‡], Zhe Wang[4,5,‡], Alexandria Do[2,3], Alexander Fuqua[3], Krista P. Balto[1], Yanning Zhang[5], Joshua S. Figueroa[1,2], Tod A. Pascal[2,3], Andrea R. Tao[1,2,3]*, Shaowei Li[1,2]*

[1]Department of Chemistry and Biochemistry, University of California, San Diego, California 92093-0309, USA

[2]Program in Materials Science and Engineering, University of California, San Diego, California 92093-0418, USA

[3]Aiiso Yufeng Li Family Department of Chemical and Nano Engineering, University of California, San Diego, California 92093-0448, USA

[4]Department of Physics and Astronomy, University of California, Irvine, California 92697-4575, USA

[5]Institute of Fundamental and Frontier Sciences, University of Electronic Science and Technology of China, Chengdu 611731, China

**ABSTRACT:** The structural and chemical properties of metal nanoparticles are often dictated by their interactions with molecular ligand shells. These interactions are highly material-specific and can vary significantly even among elements within the same group or materials with similar crystal structure. Precise characterization of ligand-metal interactions is crucial for the rational design of ligands and the functionalization of nanoparticles. In this study, we found that the ligation behavior with an *m*-terphenyl isocyanide molecule differs significantly between Au and Ag nanoparticles, with distinct ligand extraction efficiencies and size dependencies. Surface-enhanced Raman spectroscopy measurements revealed unique enhancement factors for two molecular vibrational modes between two metal surfaces, indicating different ligand binding geometries. Molecular-level characterization using scanning tunneling microscopy allowed us to directly visualize these variations between Ag and Au surfaces, which we assign as two distinct binding mechanisms. This molecular-scale visualization provides clear insights into the different ligand-metal interactions, as well as the chemical behavior and spectroscopic characteristics of isocyanide-functionalized nanoparticles.

## INTRODUCTION

Interfacial interaction between molecular ligands and metal surfaces is critical in shaping the chemical and physical properties of both the adsorbates and adsorbents.[1,2] It has been effectively employed as a key tool to engineer materials or structures with specific functions.[3,4] In nanoscience and nanotechnology, the importance of ligand-surface interaction is further amplified due to the high surface-to-volume ratio of the nanoparticles (NPs) and the frequent need for capping ligands in their preparation.[5,6] Research has demonstrated that even a minor modification in the capping ligands' structure or their surface coverage can significantly influence the NPs' properties.[7,8] Additionally, even the same ligand often interacts with different nanomaterials uniquely, due to the variation in ligand-surface interactions.[9] Consequently, precise characterization and control of ligand-surface interaction offers a promising route toward the on-demand functionalization of nanostructures.

Metal NPs exhibit unique electronic[10], optical[11] and catalytic[12] properties, making them valuable in applications for sensing[13], bioimaging[14], drug delivery[15] and heterogeneous catalysis.[16] The shape, size and stability of these NPs are usually governed by capping ligands that adsorb to the metal surface during synthesis[17] or after post-synthetic ligand exchange[18], leading to a strong dependence of NP properties on metal-ligand interactions. Surface-enhanced Raman spectroscopy (SERS)[19,20], infrared (IR) spectroscopy[21,22], surface-enhanced infrared absorption (SEIRA) spectroscopy[23,24], and nuclear magnetic resonance (NMR) spectroscopy[25,26] have provided crucial insights into the surface adsorption, coverage, and composition of these ligands. However, these ensemble-level techniques often face challenges such as signal convolution with matrix (as with NMR spectroscopy) or averaging of the spatially inhomogeneous signals (as with SERS, IR and SEIRA spectroscopy), which can obscure data interpretation and complicate the understanding behind the nature of ligand-surface interactions at the sub-NP scale. These limitations highlight the need for molecular-scale characterization to achieve a more detailed understanding of the nanoscale ligand binding to metal surfaces.

*m*-Terphenyl isocyanides are a class of ligands known for their strong metal binding ability and distinct steric profile.[27,28] These ligands feature an aryl isocyanide (i.e., CNAr; Ar = aryl) metal-binding group whose binding to the metal surface is influenced by the steric interaction between the two additional mutually meta-positioned arenes and the metal surface. Due to this steric effect, we have previously demonstrated that the *m*-terphenyl isocyanide, CNAr[Mes2] (Ar[Mes2] = 2,6-(2,4,6-Me$_3$C$_6$H$_2$)$_2$C$_6$H$_3$),[27, 29] readily binds to Au nanospheres (AuNSs) with diameters between 5 and 50 nm, but shows minimal binding to larger NPs with lower nanocurvature.[30] Given this distinct curvature threshold to ligand binding, we report here that CNAr[Mes2] binds indiscriminately to Ag nanospheres (AgNSs) of all sizes. This macroscopic contrast between Au and Ag NPs is a reflection of the unique properties, topologies and adsorption profiles of metal surfaces of differing composition. Accordingly, to provide a molecular-scale understanding of these differences, we performed microscopic investigations using a scanning tunneling microscope (STM) on CNAr[Mes2] adsorbed on Au(111) and Ag(111) surfaces. The results correlate the differences in ligand-NP interactions to the distinct adsorption structures of CNAr[Mes2] on Au and Ag surfaces. Together with first-principles electronic structure calculations which elucidate the atomic binding morphology, these findings enable a clear and coherent picture of how CNAr[Mes2] interacts differently with Au and Ag surfaces at the molecular level.

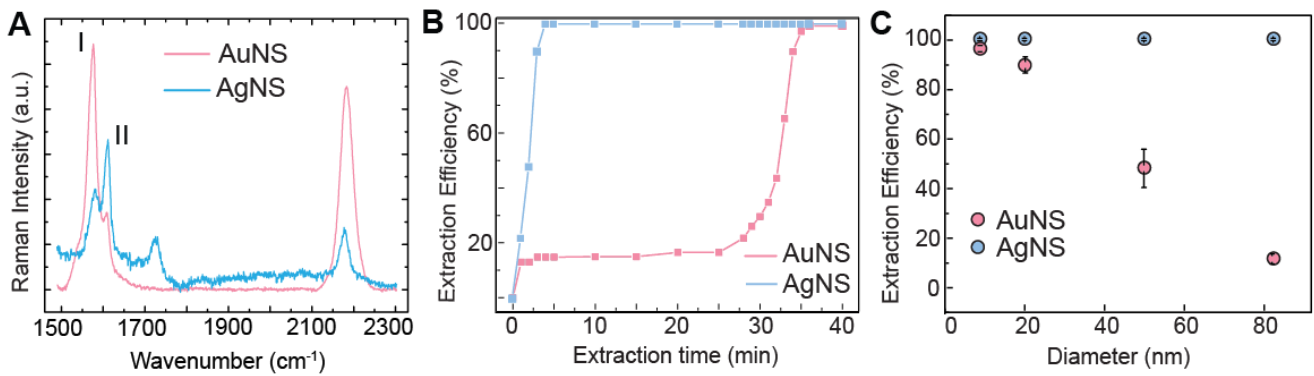

**Figure 1.** Solution-phase measurements reveal distinct binding behaviors of CNAr[Mes2] to AuNSs and AgNSs. (A) SERS spectra for AuNSs and AgNSs after LEPT with CNAr[Mes2]. (B) Plot of the extraction efficiencies of 50nm AuNSs and AgNSs extracted with CNAr[Mes2] in chloroform with respect to the extraction time. (C) Plot comparing the extraction efficiencies for AuNSs vs. AgNSs of varied sizes.

## RESULTS AND DISCUSSION

### Comparative Study on the Ligation of AuNSs and AgNSs with CNAr[Mes2]

To compare the binding behaviors of CNAr[Mes2] on citrate-stabilized pseudospherical AuNSs and AgNSs, we first performed ligand exchange in solution via phase transfer (LEPT) of citrate-stabilized AuNSs (56 ± 6 nm in diameter) and AgNSs (50 ± 4 nm in diameter) (Figure S1C) with CNAr[Mes2] between immiscible aqueous and chloroform phases (see Supporting Information for details). LEPT reveals the tendency of the binding of CNAr[Mes2] to AuNSs/AgNSs. Binding of CNAr[Mes2] to AuNSs/AgNSs was confirmed through the $\nu$(CN) stretching vibration of isocyanide at ~2180 cm$^{-1}$ with SERS, a characteristic fingerprint of the metal-bound isocyanide (Figure 1A).[19, 20, 31-35] Notably, the relative intensity between the other two vibrational modes (denoted as I and II in Figure 1A) of CNAr[Mes2] differs significantly between AuNSs and AgNSs, suggesting variation of the SERS enhancement factor due to the different molecular adsorption geometries with respect to the NP surface.[35, 36] LEPT efficiency was determined based on raffinate optical density obtained from extinction measurements (Figure S2 and section 1.4 in Supporting Information). Figure 1B presents LEPT efficiencies of the 50 nm AuNSs and AgNSs with respect to exchange time, showing a sigmoidal behavior for AuNSs that gradually increases before plateauing. This finding is consistent with previous observations that the steric interaction between CNAr[Mes2] and Au surfaces leads to a size selectivity of NPs.[30] In contrast, AgNSs are completely transported to chloroform within just 5 minutes. To further understand this difference in CNAr[Mes2] binding to Au versus Ag, we measured the LEPT efficiencies for AuNSs and AgNSs with varying diameters. Unlike AuNSs, whose extraction efficiency decreases with the diameter, AgNSs exhibit relatively high extraction efficiencies for all sizes, indicating indiscriminate CNAr[Mes2] binding to the Ag surface (Figure 1C). This large difference in LEPT efficiency indicates that CNAr[Mes2] binds more favorably to Ag than to Au NP surfaces, potentially via a different binding mechanism given the rapid ligand exchange kinetics.

### Molecular-Scale Structural Analysis with STM

To understand the difference between the ligation of AuNSs/AgNSs with CNAr[Mes2], we used STM to directly image the adsorption structure of CNAr[Mes2] on Au(111) and Ag(111) surfaces. Sample preparation procedures were previously described[37] and are detailed in section 1.5 of Supporting Information. Figure 2A shows two CNAr[Mes2] positioned side by side, straddling a step edge on Au(111). This adsorption structure is consistent with our previous report indicating that CNAr[Mes2] energetically favors the convex sites on Au(111) due to the minimum steric interference from the vicinal Au surface atoms.[37] In contrast, while CNAr[Mes2] also tends to populate the step edges on Ag(111) (Figure S3B), its appearance (Figure 2B) is significantly different from that on Au(111). Firstly, on Au(111), individual CNAr[Mes2] appears as a single, intact entity. While on Ag(111), it appears as two overlapping compartments with a larger crescent feature on top of another smaller entity. Secondly, CNAr[Mes2] covers across the Au(111) step edge but, in contrast, displays an elongated contour along the step edge of Ag(111). Further differentiation between the adsorption behaviors on Au(111) and Ag(111) was achieved in the high-resolution topographic images acquired using molecule-functionalized STM tips[38-40] (Figure 2C, D). Figure 2C clearly shows that CNAr$^{Mes2}$ spans across the Au(111) step edge, with the central aryl ring (highlighted by the bright white oval in Figure 2C) positioned directly above the step edge, and two side mesityl groups situating on the upper and lower Au layers respectively. In contrast, Figure 2D distinctly resolves the central (labeled with 1) and both of the side aryl rings (labeled with 2 and 3) of CNAr$^{Mes2}$, indicating that it resides in a parallel geometry to the Ag(111) step edge on top of another entity (labeled with 4), which we assigned as an Ag adatom. First-principles electronic structure calculations employing Density Functional Theory (DFT) were used to determine the stable binding geometries of CNAr$^{Mes2}$ on these metal surfaces. We found that the straddling geometry of CNAr$^{Mes2}$ on Au(111), where the isocyanide group bonds to an Au atom at the step edge (Figure 2E), is most stable. Similarly, the DFT-predicted adsorption geometry on Ag(111) aligns well with our hypothesis. The CNAr$^{Mes2}$ directly adsorbs on an Ag adatom and tilts towards the upper Ag layer by 28.93º relative to the (111) direction (Figure 2F and Figure S4A), which agrees with the experimental observation in Figure 2D. This adatom adsorption effectively releases the steric pressure in the side aryl rings, leading to a more stable binding compared to the straddling geometry (Figure S4B).

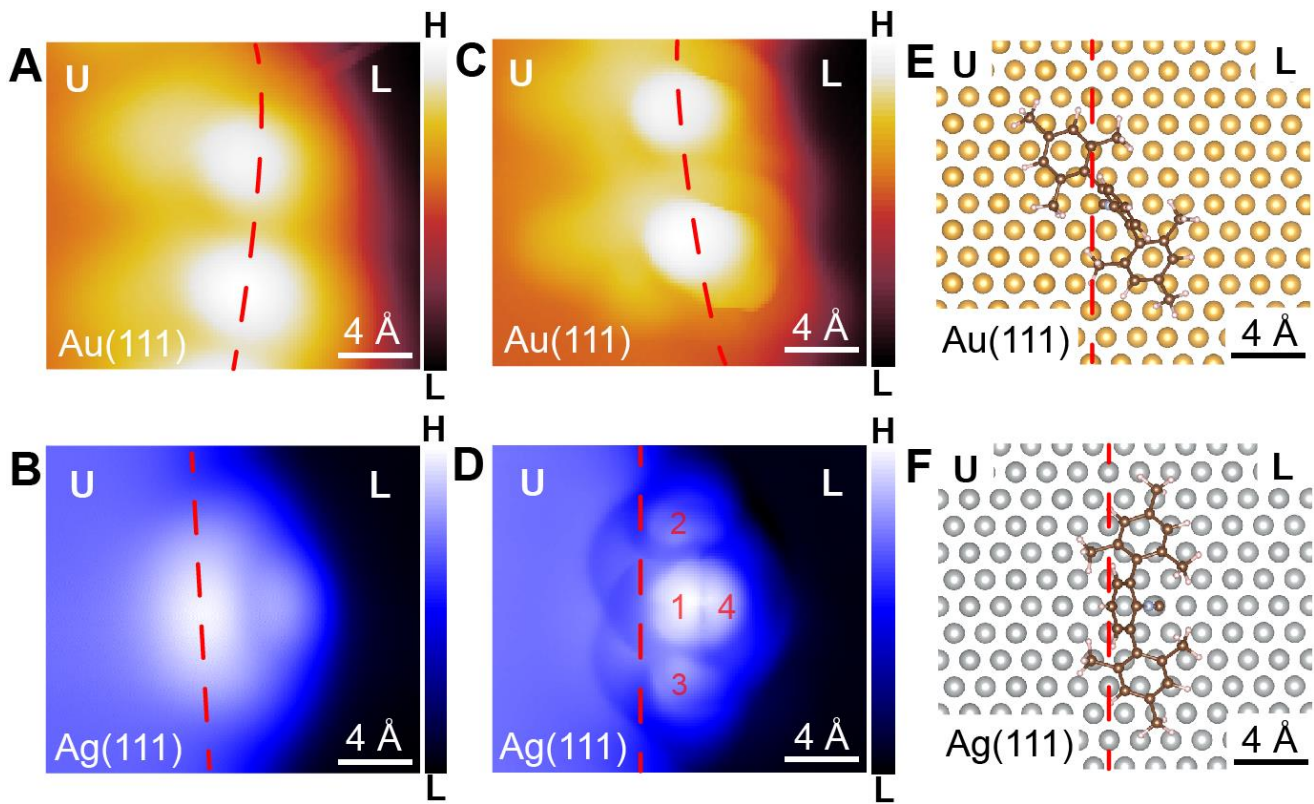

**Figure 2.** Distinct binding of CNAr$^{Mes2}$ to Au(111) and Ag(111). Conventional (A, B) and structure-resolved (C, D) STM images of CNAr$^{Mes2}$ at the step edge on Au(111) (A, C) and Ag(111) (B, D). (E, F) The top view of DFT-calculated adsorption geometries of CNAr$^{Mes2}$ at the step edge on Au(111) (E) and Ag(111) (F). Step edges are indicated by red dashed lines. Upper (U) and lower (L) surface atomic layers are labelled for clarity. Imaging parameters were -800 mV, 50 pA (A); -1 V, 20 pA (B); -565 mV, 100 pA (C); -60 mV, 100 pA (D).

**Microscopic Insights into the Distinct Binding of CNAr$^{Mes2}$ to AuNSs and AgNSs**

The distinct adsorption structures of CNAr$^{Mes2}$ on Au(111) and Ag(111) correlate directly with the differences in the ligation behavior of CNAr$^{Mes2}$ on AuNSs and AgNSs. The STM image of CNAr$^{Mes2}$ on Ag(111) (Figure 2D) indicates the involvement of Ag adatoms for CNAr$^{Mes2}$ as a dominant binding mechanism. It is worth noting that migration and diffusion of Ag atoms at room temperature across the surface has been reported in many previous studies.[41-44] These active adatoms facilitate the rapid binding of CNAr$^{Mes2}$ to AgNSs of all sizes (Figure 1B, C), without requiring corners or edges to release the steric pressure, as observed for its binding to AuNSs.[30]

The molecular-scale structural characterization also provides a clearer understanding of the variation in relative SERS signal intensity between modes I and II of CNAr$^{Mes2}$ bound to AuNSs and AgNSs (Figure 1A). According to our DFT simulations (Figure 3A), these two modes are identified as the deformation motion of the aryl rings. As shown in Figures 3B and 3C, mode I concerns the in-plane deformation of the central aryl ring (more discussion in Figure S5 and section 3.3) while mode II involves the in-plane breathing motions of the two side aryl rings. In our SERS measurements, the NPs are closely spaced and sandwiching the CNAr$^{Mes2}$ ligands within the nanogaps generated between neighboring NPs. The significantly enhanced electromagnetic field within these gaps are predominantly polarized orthogonal to the surfaces of the metallic NPs (Figure 3D), leading to the strongest enhancement of the Raman signals where molecular vibrations are aligned with the field direction.[35, 36] Consequently, the stronger signal of mode I for CNAr[Mes2] bound to AuNSs (Figure 1A) indicates an adsorption configuration where the central aryl ring is perpendicular to the surface or parallel to the electric field (Figure 3E), consistent with the straddling adsorption of CNAr[Mes2] observed on Au(111) (Figure 2C). In contrast, the weakened mode I signal for CNAr[Mes2] bound to AgNSs (Figure 1A) implies a tilted central aryl ring (Figure 3F), which agrees with the observed geometry on Ag(111) (Figure 2D).

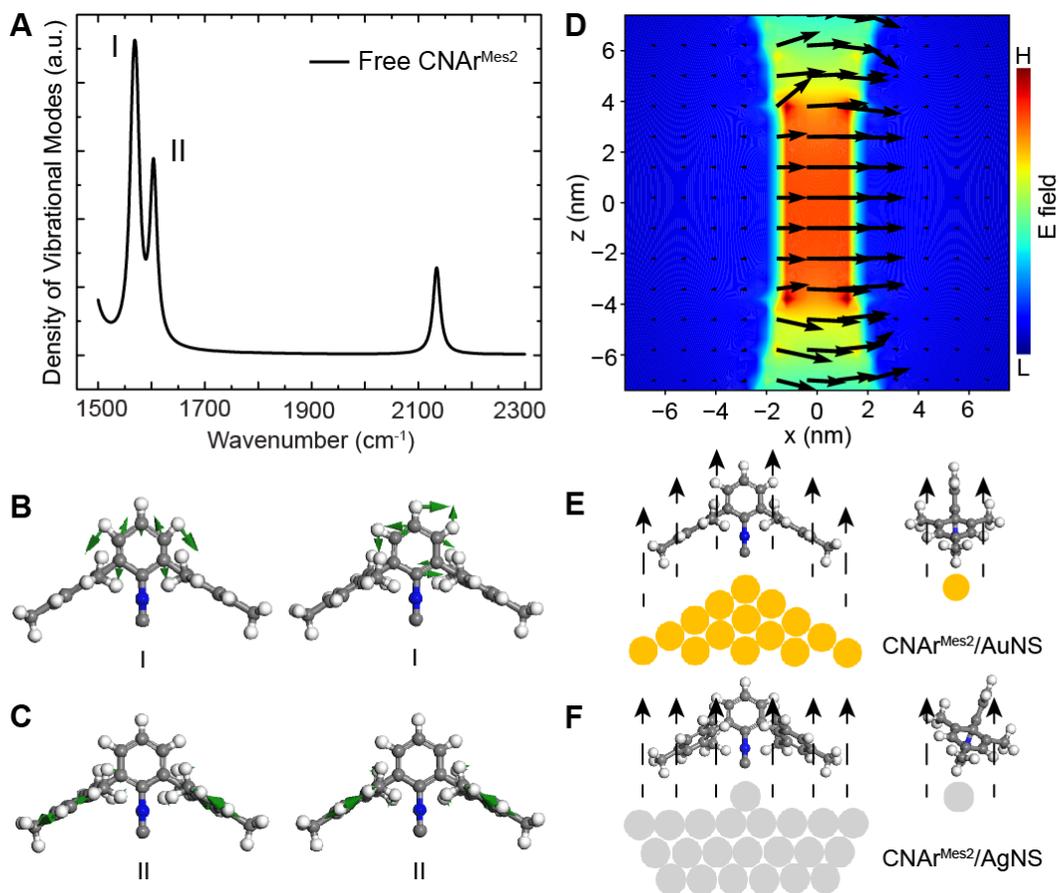

**Figure 3.** Deduced adsorption geometries of CNAr[Mes2] on AuNSs and AgNSs. (A) Calculated density of vibrational modes of free CNAr[Mes2]. (B, C) Simulated nuclear motions of a free CNAr[Mes2] for mode I (B) and II (C). The green vectors indicate the direction and amplitude of the movements. (D) Simulated near field between two 50 nm AuNSs at the plasmon resonance with a wavelength of 543 nm. The black arrows show the field directions within the nanogap. (E, F) Schematic adsorption geometries of CNAr[Mes2] on AuNSs (E) and AgNSs (F). The right panels are the side views of the CNAr[Mes2] in each configuration. Black dashed arrows indicate the direction of the electric field in the SERS measurement.

## CONCLUSIONS

By integrating the solution-phase analysis, molecular-scale characterization, and DFT calculations, we have developed a comprehensive understanding of the distinct ligation behavior of CNAr[Mes2] on AuNSs and AgNSs. On Au, CNAr[Mes2] straddles the edge and corner sites with the central aryl ring situated vertically on top of these convex surface sites (Figure 3E). On Ag, CNAr[Mes2] resides on an Ag adatom with the central aryl ring tilting towards the surface (Figure 3F).

These two distinct binding mechanisms result in different behaviors during the solution-phase ligation process, including size selectivity for AuNSs versus a rapid solvent extraction for AgNSs. This study highlights the critical role of molecular-scale characterization in unraveling the complex interactions between metal NPs and organic ligands, offering valuable insights into the interfacial chemistry of metal NPs.

## ASSOCIATED CONTENT

**Supporting Information**
Details of experimental and computational methods, and additional data to support the conclusions in the main text (PDF)
These materials are available free of charge via the Internet at http://pubs.acs.org.

## AUTHOR INFORMATION

**Corresponding Authors**

**Andrea R. Tao**
*Department of Chemistry and Biochemistry, University of California, San Diego, California 92093-0309, USA; Program in Materials Science and Engineering, University of California, San*

*Diego, California 92093-0418, USA; Aiiso Yufeng Li Family Department of Chemical and Nano Engineering, University of California, San Diego, California 92093-0448, USA;* Email: atao@eng.ucsd.edu

**Shaowei Li**
*Department of Chemistry and Biochemistry, University of California, San Diego, California 92093-0309, USA; Program in Materials Science and Engineering, University of California, San Diego, California 92093-0418, USA;* Email: shaoweili@ucsd.edu

### Author Contributions

The manuscript was written through contributions of all authors. ‡These authors contributed equally.

### Notes

The authors declare no competing financial interest.

## ACKNOWLEDGMENTS

The authors acknowledge the use of facilities and instrumentation supported by National Science Foundation (NSF) through the UC San Diego Materials Research Science and Engineering Center (UCSD MRSEC) with Grant No. DMR-2011924. This work was primarily supported by the NSF under Grant CHE-2303936 (to Shaowei Li) and DMR-2011924 (UCSD MRSEC). This work also used the Expanse supercomputer at the San Diego Supercomputing Center through allocation CSD799 from the Advanced Cyberinfrastructure Coordination Ecosystem: Services & Support (ACCESS) program, which is supported by NSF grants No. 2138259, No. 2138286, No. 2138307, No. 2137603, and No. 2138296.

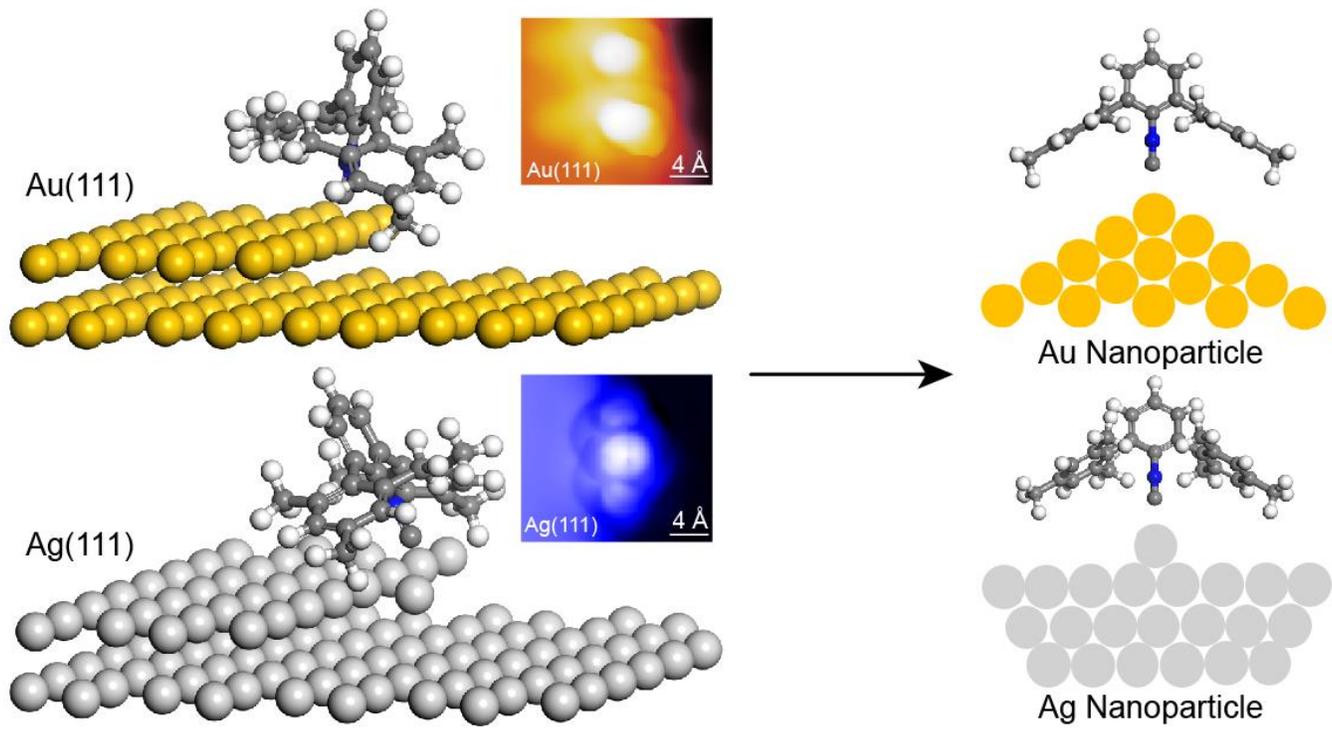

TOC Graphic